\documentstyle[12pt]{article}

\begin{document}

\title{Bloch-like oscillations induced by charge discreteness in quantum mesoscopic
rings. }
\author{J. C. Flores and E. Lazo}
\date{Universidad de Tarapac\'a, Departamento de F\'\i sica, Casilla 7-D, Arica,
Chile}
\maketitle

\baselineskip=14pt

We study the effect of charge discreteness in a quantum mesoscopic ring wih
inductance $L$. The ring is pierced by a time depending external magnetic
field. When the external magnetic flux varies uniformly, the current induced
in the ring oscillates with a frequency proportional to the charge
discreteness and the flux variation. This phenomenon is very similar to the
well known Bloch's oscillation in crystals. The similitude is related to the
charge discreteness in the charge-current representation, which plays the
same role as the constant lattice in crystals. 
\[
\]

\[
{} 
\]


05.60.G Quantum Transport Process.

07.50.E Electronics Circuits.

84.30.B Circuits Theory.

\newpage\ 

Recent advances in the development of mesoscopic physics, have allowed an
increasing degree of miniaturization and some parallel advances in
nanoelectronics. On this respect, the quantization of mesoscopic electrical
circuits appears as a natural task to undertake. In this article we discuss
the effects of a time depending magnetic flux, $\phi _{ext}(t)$, acting on a
mesoscopic ring (perfect conductor) with self-inductance $L,$ producing in
this way the equivalent to a nondissipative circuit.

From the classical point of view, the motion equation for the current can be
obtained using energy balance for this nondissipative circuit. The
electrical power $P$ transferred to a mesoscopic ring, by an external
magnetic field $B_{ext}(t),$\ is given by 
\begin{equation}
P=I\varepsilon =-I\left( \frac{d\phi _{ext}}{dt}\right) ,  \label{power1}
\end{equation}
where $I$ is the induced current; but, in the slow time variation regime,
this power is used to overcome the electromotive force in the
self-inductance $L,$ as the electric current $I$ is setting up, i.e., 
\begin{equation}
P=I\varepsilon =-I\left( L\frac{dI}{dt}\right) .  \label{power2}
\end{equation}
In this way, from (\ref{power1}) and (\ref{power2}), we obtain the
relationship:

\begin{equation}
L\frac{dI}{dt}=\left( \frac{d\phi _{ext}}{dt}\right) .  \label{motion}
\end{equation}

Because of the similitude between electric circuits and particle dynamics,
the quantization of circuits seems straightforward [1]. Nevertheless, as
pointed out by Li and Chen [2], the charge discreteness must be considered
in the quantization process. Let $q_{e}$ be the elementary charge and
consider the charge operator $\widehat{Q}$ as given by (spectral
decomposition)

\begin{equation}
\widehat{Q}=q_{e}\sum_{n}n\mid n\rangle \langle n\mid ,  \label{qcuantal}
\end{equation}
where $n$ is an integer. Following the references [2,3], and from the motion
equation (\ref{motion}), the Hamiltonian of the ring in the charge
representation is given by

\[
\widehat{H}=\frac{\hbar ^2}{2q_e^2L}\sum_n\left\{ \mid n\rangle \langle
n+1\mid +\mid n+1\rangle \langle n\mid -2\mid n\rangle \langle n\mid
\right\} 
\]

\begin{equation}
-\frac{d\phi _{ext}}{dt}q_{e}\sum_{n}n\mid n\rangle \langle n\mid .
\label{hamilton}
\end{equation}
Moreover, the current operator $\widehat{I}=\frac{1}{i\hbar }[\widehat{H},%
\widehat{Q}]$ is given explicitly by

\begin{equation}
\widehat{I}=\frac \hbar {2iLq_e}\sum_n\left\{ \mid n\rangle \langle n+1\mid
-\mid n+1\rangle \langle n\mid \right\} .
\end{equation}

The eigenstates and eigenvalues of the operator $\widehat{I}$ are easily
found. In fact, the eigenstates are

\begin{equation}  \label{currentstate}
\mid I_{k}\rangle =\sum_{n}e^{ikn}\mid n\rangle ,
\end{equation}
where the quantum number $k$ runs between $0$ and $2\pi $. The current
operator $\widehat{I}$ acting on the eigenstates (\ref{currentstate}) gives:

\begin{equation}
\widehat{I}\mid I_{k}\rangle =\frac{\hbar }{Lq_{e}}\sin (k)\mid I_{k}\rangle
,
\end{equation}
that is, the eigenvalues $I_{k}$ of the operator $\widehat{I}$ are

\begin{equation}
I_{k}=\frac{\hbar }{Lq_{e}}\sin (k),  \label{eigencurrent}
\end{equation}
which are bounded since $\left| I_{k}\right| \leq \hbar /Lq_{e}$.

As it was said before, we will study a mesoscopic ring with self-inductance $%
L$ which is pierced by a magnetic field $B_{ext}(t)$ producing a time
depending flux $\phi _{ext}(t)$. In fact, we will show that, if we begin
with one eigenstate of the current operator then, the dynamic evolution is
related to a series of states with index $k$. Explicitly, $k(t)=\phi
_{ext}(t)+k_{o},$ and then, for a homogeneously increasing magnetic flux, an
oscillating behavior of the current exists. The frequency of the
oscillations depends on the external flux variation, and is given by $\omega
=\frac{q_{e}}{\hbar }\left( \frac{d\phi _{ext}}{dt}\right) ,$which is a
constant, if $\left( \frac{d\phi _{ext}}{dt}\right) $ is a constant. We
stress the similitude between this behavior and Bloch's oscillations in
crystals under an external dc field [4-7]. In our case, it is the charge
discreteness which plays a role equivalent to the lattice constant.

In order to find the oscillations, we proceed as follows: let $\mid
k(t)\rangle $ be the state at time $t$ which is assumed as an eigenstate of
the current operator $\widehat{I}$. Let $\mid k(t+\Delta t)\rangle $ be the
state of the systems at time $t+\Delta t$. To show that this state is also
an eigenstate of the current operator, we use the first order evolution
equation

\begin{equation}
\mid k(t+\Delta t)\rangle =\mid k(t)\rangle +\frac{\Delta t}{i\hbar }%
\widehat{H}\mid k(t)\rangle .
\end{equation}
Using the commutator 
\begin{equation}
\lbrack \widehat{I},\widehat{H}]=-\frac{\hbar }{2iL}\left( \frac{d\phi _{ext}%
}{dt}\right) \sum_{n}\left\{ \mid n\rangle \langle n+1\mid +\mid n+1\rangle
\langle n\mid \right\} ,
\end{equation}
and neglecting the second order terms ($\Delta t^{2}$), we obtain

\begin{equation}
\widehat{I}\mid k(t+\Delta t)\rangle =\left( I_{k}+\frac{\Delta t}{L}\frac{%
d\phi _{ext}}{dt}\cos k\right) \mid k(t+\Delta t)\rangle .  \label{kte}
\end{equation}
That is, $\mid k(t+\Delta t)\rangle $ is an eigenstate of the current
operator with eigenvalue $\left( I_{k}+\frac{\Delta t}{L}\frac{d\phi _{ext}}{%
dt}\cos k\right) $. So, if the state of the system is initially an
eigenstate of the current operator, then it always evolves toward a state of
the current with quantum number $k(t).$ Clearly, from (\ref{kte}) and going
to the limit $\Delta t\rightarrow 0,$ we obtain the evolution equation for
the quantum number $k$ (acceleration theorem [8,9]):

\begin{equation}
\frac{dk}{dt}=\frac{q_{e}}{\hbar }\frac{d\phi _{ext}}{dt}.
\end{equation}
In this way, $k$ has a linear behavior with respect to the external flux

\begin{equation}
k(t)=\frac{q_{e}}{\hbar }\phi _{ext}(t)+k_{o},
\end{equation}
assuming that $\phi _{ext}(0)=0.$

If we consider that the magnetic flux varies uniformly with time: 
\begin{equation}
\phi _{ext}(t)=\alpha t,
\end{equation}
the quantum number $k$ becomes uniformly accelerated and then, the current $%
\langle k(t)\mid \widehat{I}\mid k(t)\rangle $ oscillates with a frequency 
\begin{equation}
\omega =\frac{q_{e}}{\hbar }\alpha .
\end{equation}

As it was said before, these oscillations in the current, and the charge,
are equivalent to Bloch's oscillations in crystals under an external dc
electric field. This analogy is very much related to charge quantization (%
\ref{qcuantal}), which plays a role similar to the constant lattice in a
crystal. Finally we note that a Hamiltonian like that described by equation (%
\ref{hamilton}), under an external dc electric field has been extensively
studied in solid state physics (tight-binding Hamiltonian). All eigenstates
are factorial localized and the spectrum is discrete [10,11]. Also, we want
to emphasize here that, as showed in [12], the discretization process
related to a Hamiltonian like (\ref{hamilton}) is not univocal.\


\begin{thebibliography}{99}
\bibitem{}  W. H. Louisell, Quantum Statistical Properties of Radiation
(Wiley, N. Y. 1973).

\bibitem{}  You-Quan Li and Bin Chen, Phys. Rev. B {\bf 53} , 4027 (1996).
Also in cond.mat/9606206 and cond-mat/9907171.

\bibitem{}  J. C. Flores, cond-mat/9908012.

\bibitem{}  F. Bloch, Z. Phys. {\bf 52}, 555 (1928).

\bibitem{}  G. Bastard, Wave Mechanics of Semiconductor Heterostructures,
Les Editions de Physique (Les Ulis, France, 1989). And references therein.

\bibitem{}  F. Rossi cond-mat/9711188. And references therein.

\bibitem{}  G. Nenciu, Rev. Mod. Phys. {\bf 63}, 91 (1991)

\bibitem{}  N. W. Ashcroft and N. D. Mermin, Solid State Physics (HRW
International editions (1988)).

\bibitem{}  O. Madelung, Introduction to Solid-State Theory
(Springer-Verlag, 1981).

\bibitem{}  M. Luban and J Luscombe, Phys. Rev. B {\bf 34}, 3676 (1986).

\bibitem{}  C. L. Roy and P. K. Mahapatra, Phys. Rev. B {\bf 25}, 1046
(1982).

\bibitem{}  J. C. Flores, J. Phys.A: Math. Gen. {\bf 26}, 4117 (1993). Also
in Physica A {\bf 192}, 665 (1993).
\end{thebibliography}
\end{document}